\documentclass[a4paper,alpha-refs]{eSpectra}

\journal{aej}

\usepackage{graphicx}
\usepackage{siunitx}
\usepackage[spanish, english]{babel} 
\usepackage{amssymb}
\usepackage{comment}

\usepackage{academicons}
\usepackage{aas-macros}
\usepackage{doi}
\usepackage{fontawesome5}
\usepackage[left]{lineno}

\definecolor{orcidlogocol}{rgb}{0.65, 0.807, 0.223}
\newcommand{\orcid}[1]{$\,$\href{https://orcid.org/#1}{\textcolor{orcidlogocol}{\faOrcid}}}
 
\title{Comparative Image Alignment for Multiple Telescopes: ALMA, IRIS, SDO}

\author[1,\authfn{1},]{Francisco J. Ordoñez Araujo}
\author[2,3,\authfn{2}]{Juan Camilo Guevara Gómez}
\author[1,\authfn{3}]{Benjamín Calvo Mozo}

\affil[1]{Observatorio Astronómico Nacional, Universidad Nacional de Colombia, Bogotá D.C., Colombia}
\affil[2]{Group Research \& Development DNV AS, Høvik, Norway}
\affil[3]{Rosseland Centre for Solar Physics, University of Oslo, Postboks 1029 Blindern, 0315 Oslo, Norway}

\authnote{\authfn{1} Master of Science Student - Astronomy (fjordoneza@unal.edu.co)}
\authnote{\authfn{2}Researcher (juan.camilo.guevara.gomez@dnv.com)}
\authnote{\authfn{3}Associate professor  (bcalvom@unal.edu.co )}

\papercat{Scientific Article}

\runningauthor{Sánchez Gonzalez et al.}

\jvolume{2}
\jnumber{1}
\jyear{2024}
\begin{document}
\begin{frontmatter}
\maketitle

\selectlanguage{english}
\begin{abstract}
\justifying
This paper introduces a method for aligning solar observations from different telescopes. We utilized helioprojective coordinates from the Solar Dynamics Observatory (SDO) as a reference to align images from ALMA and IRIS. The alignment is based on correlation analysis, employing both the Pearson Correlation Coefficient (PCC) and the Structural Similarity Index (SSIM) for assessing data correlations. PCC was preferred for its effectiveness with diverse datasets and its computational efficiency compared to SSIM. The alignment demonstrated less than 1.0\,arcsec variation in average centers between the two methods, ensuring consistent results. We illustrate this method using ALMA observations, labeled D06 in the Solar ALMA Science Archive (SALSA), as a case study.
\end{abstract}


\selectlanguage{spanish} 
\begin{abstract}
\justifying
Este trabajo introduce un método para alinear observaciones solares provenientes de telescopios distintos. Utilizamos las coordenadas helioproyectivas del Observatorio de Dinámica Solar (SDO) como referencia para alinear las imágenes de ALMA e IRIS. El proceso de alineación se basa en un análisis de correlación, utilizando tanto el Coeficiente de Correlación de Pearson (PCC) como el Índice de Similitud Estructural (SSIM) para estimar las correlaciones entre los conjuntos de datos. Se seleccionó el PCC como el método preferido debido a su fiabilidad en casos de conjuntos de datos muy diferentes y sus ventajas computacionales significativas en comparación con el SSIM. Los resultados de la alineación demuestran una variación de menos de 1 segundo de arco entre los centros promedio obtenidos por ambos métodos, asegurando la consistencia en los resultados finales. Como caso ilustrativo, hemos utilizado observaciones de ALMA identificadas como D06 en el Solar ALMA Science Archive (SALSA) para explicar y mostrar el método.
\end{abstract}

\begin{skeywords}
Alineación de Imágenes Solares--  Atmósfera Solar-- Coordenadas Helioproyectivas-- PCC-- SSIM.
\end{skeywords}
\end{frontmatter}

\selectlanguage{english}

\section{Introduction}

The study of the solar atmosphere has garnered significant importance due to a variety of factors. These encompass fundamental inquiries, the best known being the elucidation of the origins of the remarkably high temperatures observed in the solar corona, commonly referred to as the ``coronal heating problem''. Additionally, it encompasses the examination of space weather and the impact of solar atmospheric phenomena, such as solar flares and coronal mass ejections, on terrestrial weather conditions.

In the pursuit of understanding the solar atmosphere, spectroscopy has played a pivotal role by enabling the analysis of its composition and properties. Semi-empirical models, exemplified by the work of \cite{vernazza_structure_1981}, have offered initial insights into the distinctive emission lines characterizing each atmospheric layer. For instance, the photosphere is recognized for its prominent H-alpha emission, the chromosphere for its strong radio emissions, and the transition region and corona exhibit intense ultraviolet emissions.

This knowledge of characteristic emissions from various solar atmospheric layers has driven the development of specialized instruments, including the  Interface Region Imaging Spectrograph (IRIS; \cite{depontieu_interface_2014_IRIS}), the Solar Dynamics Observatory (SDO; \cite{SDO_pesnell}), and the Atacama Large Millimeter/submillimeter Array (ALMA; \cite{wootten_atacama_2009}). These instruments are tailored to observe precise wavelengths corresponding to emissions coming from distinct layers, significantly enhancing our ability to investigate the underlying physical phenomena across the solar atmosphere.

However, studying the solar atmosphere requires observing its ever-changing physical conditions. To compare different atmospheric layers and understand diverse physical phenomena, data processing becomes essential. Research articles often provide limited information on data pre-processing, as they focus more on the specific research goals of the authors. Hence, in this paper, we introduce a pre-processing method that is crucial for analyzing data captured by IRIS, SDO, and ALMA in the study of the solar atmosphere’s dynamics. We particularly emphasize an alignment technique that enables the synchronization of images captured by these telescopes, ensuring more accurate comparative analysis.

\section{Method}
\subsection{Observational data}

In this study, we utilized observational data from various telescopes including the Helioseismic and Magnetic Image (HMI; \cite{scherrer_helioseismic_2012}), the Atmospheric Imaging Assembly (AIA; \cite{lemen_atmospheric_2012}) on the SDO, IRIS, and ALMA in Band 3. Specifically, we analyzed brightness temperature data for observation D06 (project ID: 2017.1.00653.S), sourced from the Solar ALMA Science Archive  (SALSA; \cite{SALSA_guia}). This region, located in the Southern Hemisphere, was observed from April 12, 2018, from 15:52:28 to 16:24:41 UTC. The observation targeted helioprojective coordinates ($T_x$=-128, $T_y$=-400), as reported by the \href{http://sdc.uio.no/salsa/}{SALSA} database.

The observation time for the D06 region totaled 32\,min, split into three consecutive scans of 9.9, 10, and 7.0\,min, interspersed with 2.5 minute gaps. The FoV of the observation correspond to a circle with radius 37.95\,arcsec. The images are characterized by a pixel size of 0.3\,arcsec. The ALMA synthesized beam which represents the spatial resolution varies over time, with average values of 1.77\,arcsec for the minor axis and 2.55\,arcsec for the major axis. Furthermore, the  beam displays an average inclination of 77.5 degrees with respect to solar north. The cadence of the ALMA observation corresponds to 1.0 second, resulting in a total of 1608 images distributed across the three scans. The ALMA images exhibit an angle of inclination of 26 degrees counterclockwise with respect to solar north. Additionally,  the heliocentric angle ($\mu = \cos{\theta}$)  of the region is 0.90 with an average brightness temperature of $7689 \pm 661$\,K.  

The IRIS observation utilized in this study spanned 9.6\,hours, starting on April 12, 2018, at 13:09 and concluding at 22:50. The reference helioprojective coordinates at the initiation of IRIS observation were $T_x, T_y = -118.181, -388.526$. During this period, IRIS conducted observations concurrently with ALMA in the D06 and D26 datasets in Band 3, and D27 and D09 datasets in Band 6, as part of project 2017.1.00653.S, (refer to the SALSA database for details).  The FoV of the IRIS observation is 128.5\,arcsec$\times$122.0\,arcsec. The images are characterized by a pixel size of 0.1663\,arcsec. The orientation of IRIS images aligns with solar north. Throughout the observation period, IRIS collected a total of 980 images at a cadence of 36.696 seconds. In the far-ultraviolet channel (1332-1358\,Å and 1390-1406\,Å), IRIS achieved a spectral resolution of 40\,mÅ, while in the near-ultraviolet channel (2785-2835\,Å), it achieved a spectral resolution of 80\,mÅ. The effective spatial resolution of IRIS during the observation ranged between 0.33 and 0.4\,arcsec.

HMI offers magnetograms by observing the iron line Fe I (6173\,Å) at photospheric altitudes across the solar disk. During the observation window of interest, HMI acquired a set of 43 images at a capture cadence of 45 seconds per frame with a pixel size of 0.4\,arcsec. Additionally, images of the mid-to-high photosphere and high chromosphere regions were obtained through the AIA/1600\,Å, AIA/1700\,Å, and AIA/304\,Å channels, featuring a pixel size of 0.6 arcsec and capture cadences of 24 seconds for AIA/1600\,Å and AIA/1700\,Å, and 12 seconds for AIA/304\,Å. It is noteworthy that both HMI and AIA images exhibit a 180 degrees inclination with respect to the solar north.

\subsection{Alignment process} \label{chap_2:seccion_2.3}

In this work, the co-alignment of images from SDO, IRIS, and ALMA telescopes was achieved by analyzing their similarity. The Pearson correlation coefficient (PCC) was used to study the linear correlation between the images \cite{yeager_libguides_nodate}, while the Structural Similarity Index (SSIM) was used for a more detailed analysis \cite{wang_image_2004_SSIM}. 

The PCC statistic ranges from -1 to +1. A value of -1 indicates a perfect linear anti-correlation between the variables, meaning that as one variable increases, the other decreases. A value of +1 indicates a perfect linear correlation between the variables, signifying that as one variable increases, the other also increases. A value of 0 suggests no linear correlation between the variables.

The SSIM metric estimates for an image pair fall within the range of values between -1 and +1. A value of -1 signifies that the compared images are entirely dissimilar, while a value of +1 indicates that the images are either identical or highly similar. It's worth noting that in some libraries, like the one used in this work (\href{https://scikit-image.org/docs/stable/auto_examples/transform/plot_ssim.html}{Scikit-image}), the SSIM metric is adjusted to a range between 0 and +1 for convenience.


The alignment process was divided into three distinct stages. In the first stage,  as detailed in Section \ref{data_preparation}, data was downloaded, and images were rotated to ensure their alignment with the solar north. 
The second step, described in Section \ref{alma-IRIS_correlation_process}, focused on identifying the pixel with the maximum correlation between IRIS and ALMA within the IRIS data. The third and final stage, outlined in Section \ref{Alignment_of_IRIS_images_with_SDO}, was dedicated to the precise alignment of IRIS and SDO images to determine the helioprojective coordinates of the pixel exhibiting the maximum correlation between ALMA and IRIS, as estimated in the second step of the alignment process.


\subsection{Data acquisition and  preparation}
\label{data_preparation}

The alignment method presented in this work begins with the precise download and selection of data. Aligning images captured by different telescopes is only possible when they observe the same solar region during a shared observation time window. The SALSA database stores information about regions of the solar chromosphere observed by ALMA in bands 3 and 6, allowing us to identify regions that have been co-observed with IRIS and SDO.

In this study, we have chosen the dataset D06 from SALSA, which is a quiet sun region that has been co-observed by the aforementioned telescopes. The data corresponding to the co-observation of IRIS with ALMA in region D06 were downloaded from the \href{https://iris.lmsal.com/search/}{IRIS database}. 

The SDO data were acquired using the \textit{Sunpy} library and the \href{http://jsoc.stanford.edu/}{Joint Science Operations Center (JSOC)} database.  
The SDO data were downloaded and cropped by tracking the helioprojective coordinates corresponding to the initial time provided in the SALSA database for D06, with $T_x$ and $T_y$ values of -128 and -400, respectively. This option was enabled in the JSOC database download settings. The cropping was performed such that the resulting SDO images had a radius 1.8 times larger than the FoV radius of the ALMA data.

The SDO and ALMA images were rotated using the \textit{CROTA2} \cite[See Table 1,][]{brown_working_nodate}
and \textit{SOLAR\_P} \cite[See Table A.2,][]{SALSA_guia}
values found in the header, applying corrections of 180 degrees and 26 degrees, respectively, to ensure that the north of the images aligned with the solar north. 
In contrast, the IRIS images did not require rotation, as their orientation already matched the solar north. After these adjustments, all images were oriented to the solar north, ensuring consistent alignment throughout the dataset.

The data used in the co-alignment correspond to  IRIS telescope at wavelengths Mg II wing 2832\,Å, Mg II h/k 2796\,Å, C II 1330\,Å, and Si IV 1400\,Å, SDO/AIA at wavelengths 304\,Å, 1600\,Å, 1700\,Å,  and  SDO/HMI. In the IRIS data cube, frames corresponding to the observation window of region D06 were selected.

\subsection{Pixel-Level correlation localization between ALMA and IRIS}
\label{alma-IRIS_correlation_process}

To estimate the pixel of maximum correlation between the IRIS and ALMA images, a detailed process was followed. Initially, the IRIS images were re-scaled. This re-scaling used the ratio between the pixel sizes of IRIS and ALMA as a factor, ensuring that observations from both telescopes had the same pixel size (0.1663/0.3), resulting in a uniform pixel size of 0.3\,arcsec for both sets of images. During this process, the pixel dimensions of the data, originally spanning a FoV of 128.5\,arcsec$\times$122.0\,arcsec, were altered. The dimensions shifted from an original size of 774$\times$735\,pixel to a resized dimension of 429$\times$408\,pixel, reflecting the scaling adjustments necessary to achieve uniformity across both telescopic observations.

In the next phase, the IRIS frames that were co-observed alongside ALMA were specifically selected for alignment. Both sets of images then underwent an averaging and interpolation process to standardize their intensity scales. 
This step was crucial to ensure consistency between the IRIS and ALMA data, with the intensity scale ranging from 0 to +1. 
This phase involved creating two synchronized images, representing concurrent observations from IRIS and ALMA. These images were prepared for correlation analysis in the subsequent step to facilitate alignment, an essential preparatory phase for the upcoming stages of the process.

To analyze the relationship between the IRIS and ALMA images, we conducted a detailed correlation analysis. 
This involved a unique mapping process, given the difference in the FoV sizes of the two images, with the IRIS FoV being larger than that of ALMA. Specifically, the smaller ALMA image was initially aligned with one corner of the larger IRIS image. From there, we employed a moving window approach, systematically shifting the ALMA image pixel-by-pixel across the entire area of the IRIS image.  At each new position, the center pixel of the ALMA image corresponded to a specific location on the IRIS image. For every such alignment, we calculated the Pearson Correlation Coefficient (PCC) and Structural Similarity Index (SSIM), associating these values with the IRIS image location corresponding to the ALMA image's central pixel. This comprehensive scanning process effectively created a matrix of PCC and SSIM values that matched the dimensions of the IRIS image, reflecting the correlation metrics at every possible alignment of the two images within the IRIS FoV.

Following the comprehensive scanning and correlation analysis, our next objective was to pinpoint the precise coordinates in the averaged and interpolated IRIS matrix where the highest correlation with the ALMA images occurred. These specific coordinates indicate the optimal alignment position for each IRIS filter image with the ALMA image.

We replicated this procedure for every IRIS filter. To refine our results, we averaged the coordinates identified across all filters as the points of maximum correlation. This averaged coordinate location was then regarded as the definitive alignment point within the IRIS array for the images from both IRIS and ALMA telescopes.

The final stage in our alignment process was crucial: determining the helioprojective coordinates of the pixel exhibiting the peak correlation value. Successfully identifying these coordinates was imperative for achieving an accurate and meaningful alignment, thereby ensuring a precise correlation between the images captured by the two telescopes.

\begin{figure*}[htb!]
   \centering
    \includegraphics[width=\textwidth]{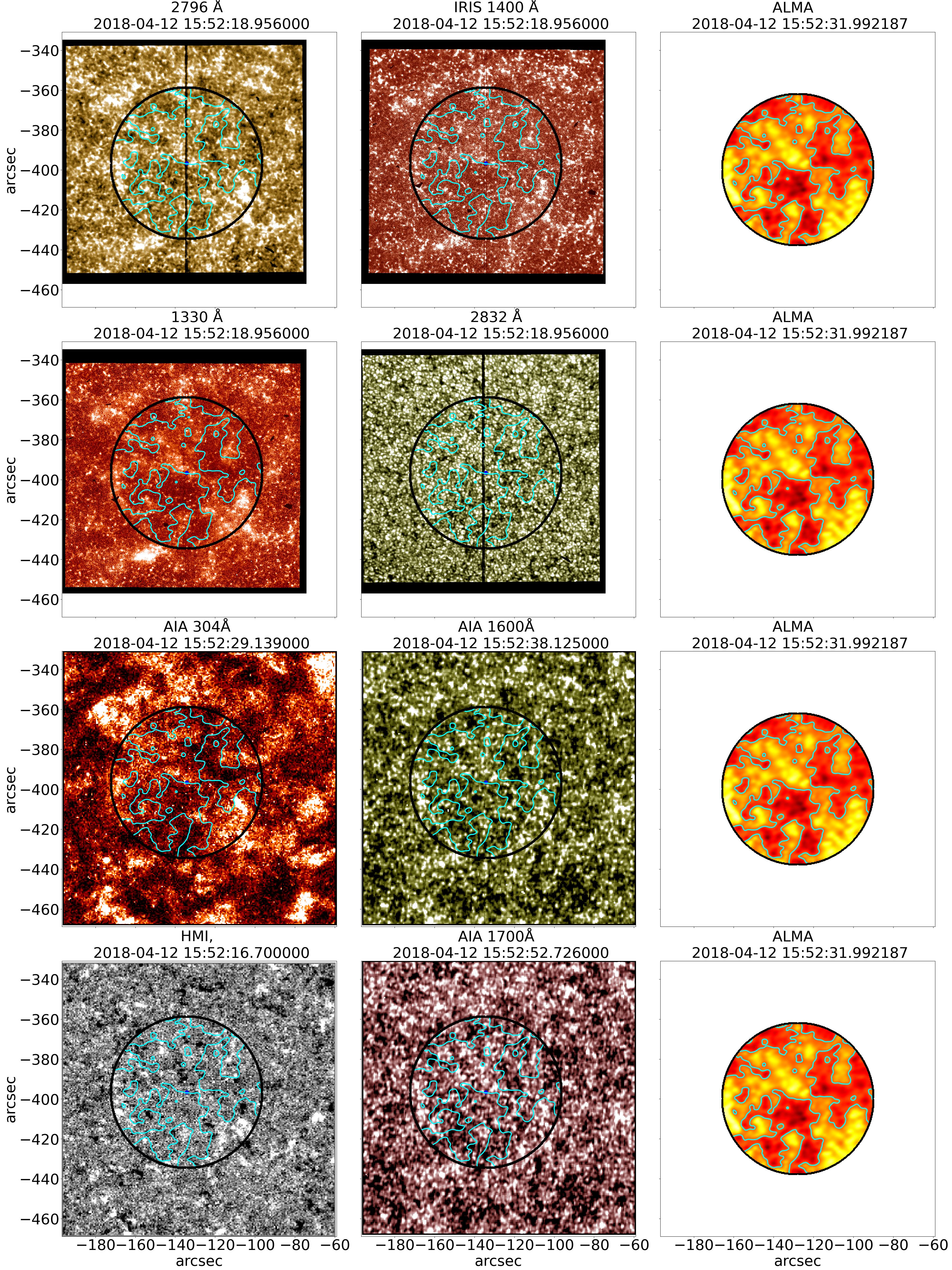}
    \caption{
    Result of the alignment  ALMA Band 3, SDO,  and IRIS for region D06 of the \href{http://sdc.uio.no/salsa/}{SALSA} database, corresponding to the observation of April 12, 2018 at 12:15 UTC. The black circle represents the ALMA FoV, which is centered on the helioprojective coordinates $T_x,$ $ T_y$ = -134.21 -396.86, which were estimated during the alignment process. The regions of the chromosphere with a brightness temperature of  7500 K in the ALMA images are delimited by cyan contours.}
    \label{chap_2:fig_result_IRIS_and_SDO_aling}
\end{figure*}

\subsection{Alignment of IRIS images with SDO}
\label{Alignment_of_IRIS_images_with_SDO}

The alignment process between the IRIS and SDO images, essential for ensuring that ALMA aligns with SDO through IRIS, involved several meticulous steps. Initially, the IRIS images were cropped, centering each frame on the pixel identified as having the maximum correlation with the ALMA images. This strategic alignment of IRIS with ALMA was crucial, as it ensured that, by aligning IRIS with SDO in subsequent steps, ALMA would also align with SDO by transitivity. Additionally, the cropping expanded the field of view by approximately 10 arcsec to enhance computational efficiency. 

Due to IRIS's use of four different intensity filters, specific standardization steps were necessary for accurate comparison with SDO images. The IRIS data were normalized by their standard deviation, ensuring the filters with higher intensities contributed appropriately to the averaged data used in later stages.

The cropped IRIS images were rescaled using a scaling factor derived from the ratio between the pixel size in arcsec of IRIS and that of SDO (0.1663/0.6), ensuring both images had the same pixel size (0.6\,arcsec). During this process, the pixel dimensions of the cropped IRIS data, originally spanning a FoV of 38.25\,arcsec $\times$ 38.25\,arcsec, were adjusted. The dimensions transitioned from their original size of 230$\times$230\,pixels to a resized dimension of 160$\times$160\,pixels. This adjustment reflects the scaling changes required to achieve uniformity in both observations.

Next, simultaneous observations of SDO images and cropped IRIS frames (across all four filters) with ALMA were selected. With uniform cadences and pixel sizes across the filters, an average of all IRIS filter images was computed to create a representative IRIS data cube. This IRIS cube was then aligned with the SDO images. 
Both data sets were averaged and interpolated to standardize intensity scales, facilitating the correlation analysis.

The correlation analysis involved mapping the cropped IRIS images with the SDO images. As detailed in Section \ref{alma-IRIS_correlation_process}, for each mapping, the Pearson Correlation Coefficient (PCC) and the Structural Similarity Index (SSIM) were estimated and stored in a matrix corresponding to the dimensions of the SDO matrix.

The process also involved determining the position in coordinates of pixels in the averaged and interpolated SDO matrix, where the maximum correlation between the IRIS images with SDO was found. This position corresponds to the pixel where both images are aligned for a given SDO filter. The helioprojective coordinates of the pixel where the maximum correlation was found between the IRIS and SDO images were estimated using the information in the SDO data headers. This step was repeated for each SDO filter, averaging the helioprojective coordinates of the pixels with the highest correlation, thereby aligning the IRIS images using SDO coordinates as a reference. Consequently, the IRIS images were aligned with the SDO images. By extension, as in the second step, IRIS's alignment with ALMA implied the alignment of ALMA images with SDO as well.

To manage computational efficiency in this complex alignment process, we focused on optimizing the search for maximum correlations. The decision to crop SDO data to a FoV 1.8 times that of ALMA's Band 3 observations was made to limit the spatial range for potential correlations. In the initial step of Section \ref{Alignment_of_IRIS_images_with_SDO}, cropping the IRIS data created a FoV marginally larger than ALMA's. This approach served a dual purpose: it reduced the number of correlation calculations and ensured that the images to be aligned (IRIS and ALMA) were smaller in arcsecond dimensions compared to the SDO data, which acted as the reference in the alignment process.

The data acquisition and image alignment process was implemented in Python. The documentation of the programs used as well as the script used to align the images is available on GitHub \cite{gibhub_reporsitory_alignment} together with a guide for the use of the program. Also, you find more information about the proposed method in the second chapter, the master's thesis \cite{araujo_dynamics_2024}.

\begin{table*}[!htp]
\centering
\begin{tabular}{ccccccc}
\toprule
\toprule
\multicolumn{7}{c}{ALMA Alignment with IRIS} \\ 
\midrule
\multicolumn{1}{c}{IRIS Filters(Å)} & \multicolumn{3}{c}{Pearson} & \multicolumn{3}{c}{SSIM} \\ 
\cmidrule(lr){2-4} \cmidrule(lr){5-7}
& pixel x & pixel y & r & pixel x & pixel y & r \\
\midrule
1330.0 & 371.51 & 360.69 & 0.72 & 371.51 & 362.49 & 0.26 \\
1400.0 & 373.31 & 362.49 & 0.56 & 373.31 & 362.49 & 0.14 \\
2796.0 & 376.92 & 362.49 & 0.52 & 375.11 & 369.70 & 0.15 \\

\textcolor{red}{2832.0} & \textcolor{red}{375.11} & \textcolor{red}{367.90} & \textcolor{red}{0.08} & \textcolor{red}{472.50} & \textcolor{red}{449.05} & \textcolor{red}{0.03}\\
Mean   & 373.91 & 361.89 & 0.60 & 373.31 & 364.89 & 0.18 \\
\bottomrule
\end{tabular}
\caption{Results of identifying the pixel with the maximum correlation between IRIS and ALMA images for region D06 from the \href{http://sdc.uio.no/salsa/}{SALSA} database, corresponding to the observation of April 12, 2018 at 12:15 UTC. The red data were not considered in the average.}
\label{chap_2:result_IRIS_and_ALMA_align}
\end{table*}

\begin{table*}[!htp]
\centering
\begin{tabular}{ccccccc}
\toprule
\toprule
\multicolumn{7}{c}{IRIS Alignment with the SDO} \\ 
\midrule
\multicolumn{1}{c}{SDO Filters(Å)} & \multicolumn{3}{c}{Pearson} & \multicolumn{3}{c}{SSIM} \\ 
\cmidrule(lr){2-4} \cmidrule(lr){5-7}
& Tx & Ty & r & Tx & Ty & r \\
\midrule
304 & -136.66 & -398.02 & 0.46 & -137.21 & -398.02 & 0.09 \\
HMI & -134.21 & -395.69 & 0.35 & -133.70 & -395.69 & 0.11 \\
1600 & -133.16 & -397.25 & 0.54 & -133.16 & -396.25 & 0.24 \\
1700 & -133.12 & -396.47 & 0.52 & -133.12 & -396.47 & 0.23 \\
Mean & -134.21 & -396.86 & 0.47 & -134.30 & -396.61 & 0.17 \\
\bottomrule
\end{tabular}
\caption{Results of the alignment of the IRIS images with the SDO images for region D06 from the \href{http://sdc.uio.no/salsa/}{SALSA} database, corresponding to the observation of April 12, 2018 at 12:15 UTC.}
\label{chap_2:result_IRIS_and_SDO_align}
\end{table*}

\section{Results and analysis}

The result of applying the process of finding maximum correlation pixels between the IRIS and ALMA images,  detailed in Section \ref{alma-IRIS_correlation_process}, is presented in Table \ref{chap_2:result_IRIS_and_ALMA_align}.
This table displays the pixel coordinates (pixel x, pixel y) within the IRIS matrix where the maximum correlation has been found between ALMA data and each of the IRIS filters, using the SSIM and PCC methods. Additionally, the table includes the correlation value (r) obtained in the fit.


In general, the pixel positions obtained by the PCC and SSIM methods present a percentage discrepancy of less than $2\%$, see Table \ref{chap_2:result_IRIS_and_ALMA_align}.
However, in the particular case of the correlations between IRIS 2832\,Å images and ALMA images in the D06 dataset, the percentage discrepancy reaches about $26\%$. This discrepancy is due to the fact that IRIS 2832Å is observing the mid-high photosphere \cite{grubecka_height_2016}, while ALMA Band 3 in D06 is observing the mid-chromosphere \cite{wedemeyer_sun_2020_ALMA_band_3}. Therefore, the estimate of the SSIM index for this data set is unreliable. In contrast, the PCC is more reliable in this case because it focuses on estimating the linear correlation between the data, rather than comparing the brightness, contrast, and structure of two very different data sets as the SSIM method does.

In a similar study, \cite{sim_vs_pcc_2020} explored the effectiveness of the SSIM and PCC methods as criteria for measuring image similarity. \cite{sim_vs_pcc_2020}'s findings revealed that the SSIM method is not capable of accurately determining image similarity, proving effective only in cases of highly visually similar images that belong to the same scene. This is particularly evident in the case of IRIS 2832\,Å, and ALMA in region D06, where, despite representing the same moment in time, the coupling between layers is weak, as illustrated in Figure \ref{chap_2:fig_result_IRIS_and_SDO_aling}. 
However, it is possible that the SSIM method could perform better in active regions of the solar atmosphere, where layer coupling is stronger, especially when comparing images such as IRIS 2832\,Å and ALMA band 3.
Furthermore, \cite{sim_vs_pcc_2020} concludes in his research that the PCC method emerges as a more reliable tool for measuring image similarity and, in terms of computational efficiency, is quicker.


Hence, it was decided that in the computational implementation of the alignment process outlined in Section \ref{alma-IRIS_correlation_process}, the PCC method would be employed to estimate the correlations between IRIS and ALMA. This choice was made because the SSIM method incurs a 40\% higher computational cost. Furthermore, when comparing the average centers obtained by both methods, excluding the correlation between IRIS 2832\,Å and ALMA band 3, the difference between them introduces variations of less than 1.0\,arcsec in the final alignment result.




The Table \ref{chap_2:result_IRIS_and_SDO_align} presents the results obtained in the process of aligning IRIS images with SDO, as detailed in Section \ref{Alignment_of_IRIS_images_with_SDO}. It shows the helioprojective coordinates estimated for the average position with the highest correlation, identified in the second step of the alignment process. 
In particular, it is observed that for the average position of maximum correlation (x,y) = (374.21, 363.39) obtained with the PCC statistic (see Table \ref{chap_2:result_IRIS_and_ALMA_align}), the average helioprojective coordinates are $T_x$ = -134.21 and $T_y$ = -396.86. Additionally, using the SSIM metric, the coordinates are found to be $T_x$ = -134.30 and $T_y$ = -396.61. The coordinates we have determined exhibit a percentage discrepancy of $4.5\%$ and $0.8\%$ for the $T_x$ and $T_y$ components, respectively, when compared to those reported by \cite{SALSA_guia}. 



Using the helioprojective coordinates obtained through the proposed alignment method, along with the headers from SDO and IRIS data, we created Figure \ref{chap_2:fig_result_IRIS_and_SDO_aling} to visually illustrate the alignment results. In the left and center columns, we present the SDO and IRIS observations, respectively, in the specified filters as indicated in the titles.  The black circle in these columns denotes the results obtained using the proposed alignment method. In the right column for each row, we display the ALMA data,  enabling a direct visual comparison of the method's effectiveness in terms of the similarity between ALMA's thermal structure and the images in the left and center columns. The cyan contour corresponds to temperature, 7500\,K.

During program development, we conducted tests to assess direct alignment between ALMA images with SDO and IRIS images with SDO. In these tests,   the helioprojective coordinates obtained from both alignments differed by less than 0.5\%, approximately 2.0\,arcsec, which equated to a discrepancy of 6 pixels. This was also consistent with the results obtained using the method proposed in this study, where the discrepancy remained below 2.0\,arcsec. Our research revealed that the best results were achieved when aligning IRIS, ALMA, and SDO images using the method outlined in this study. Additionally, in the GitHub repository \cite{gibhub_reporsitory_alignment}, you can also find routines for aligning ALMA images with SDO and IRIS images with SDO.

\section{Conclusions}
The proposed alignment process, detailed in   Sections \ref{chap_2:seccion_2.3}, \ref{alma-IRIS_correlation_process}, and \ref{Alignment_of_IRIS_images_with_SDO},
successfully aligns ALMA and IRIS images using the helioprojective coordinates from SDO as a reference point. The percentage discrepancy between the helioprotective coordinates $T_x$ and $T_y$ obtained in our methodology and those reported by \cite{SALSA_guia} in the ALMA database for the D06 was $4.5\%$ and $0.8\%$, respectively. It is worth noting that the coordinates given in SALSA correspond to a primary, non-systematic co-alignment carried out during the image reconstruction of the datasets. This success is clearly reflected in Figure \ref{chap_2:fig_result_IRIS_and_SDO_aling}, where a clear correlation is observed between the thermal structure of ALMA data and observations from SDO and IRIS in the corresponding filters.

The alignment process employs both the PCC and the SSIM methods to estimate correlations between the datasets. PCC is chosen as the preferred method due to its reliability in cases of vastly different datasets and its significant computational advantages over SSIM. The results exhibit less than a 1.0\,arcsec variation between the average centers determined by the two methods, ensuring consistency in the final alignment results.

It is worth noting that the proposed method can also be used for aligning images from other telescopes. However, caution should be exercised when selecting the AIA filter to serve as a reference for alignment. The choice should ensure that the formation height of the emission line in the reference filter is close to the formation height of the data being aligned. This will ensure an effective and accurate correlation between the images, as the method relies on analyzing the correlation and similarity between them. It should be also relatively simple to adapt the method in the case that there were not co-observations with IRIS.

In future research endeavors, this method could serve as a foundation for the development of an advanced alignment technique that takes into account the projection effect when aligning images. Such an approach could further enhance the precision of image alignment and contribute to a deeper understanding of the small-scale solar phenomena.

\section*{Acknowledgments}
This paper makes use of the following ALMA data: ADS/JAO.ALMA\#2017.1.00653.S. ALMA is a partnership of ESO (representing its member states), NSF (USA), and NINS (Japan), together with NRC (Canada), MOST and ASIAA (Taiwan), and KASI (Republic of Korea), in co-operation with the Republic of Chile. The Joint ALMA Observatory is operated by ESO, AUI/NRAO, and NAOJ. SALSA, SALAT, and SoAP are produced and maintained by the SolarALMA project, which has received funding from the European Research Council (ERC) under the European Union's Horizon 2020 research and innovation programme (grant agreement No. 682462), and by the Research Council of Norway through its Centres of Excellence scheme, project number 262622.




\bibliography{biblio}

\begin{thebibliography}{}

\bibitem [\protect \citeauthoryear {%
Brown%
, Regnier%
, Marsh%
\BCBL {}\ \BBA {} Bewsher%
}{%
Brown%
\ \protect \BOthers {.}}{%
{\protect \APACyear {2010}}%
}]{%
brown_working_nodate}
\APACinsertmetastar {%
brown_working_nodate}%
\begin{APACrefauthors}%
Brown, D.%
, Regnier, S.%
, Marsh, M.%
\BCBL {}\ \BBA {} Bewsher, D.%
\end{APACrefauthors}%
\unskip\
\newblock
\APACrefYearMonthDay{2010}{}{}.
\newblock
{\BBOQ}\APACrefatitle {Working with data from the {Solar} {Dynamics} {Observatory}} {Working with data from the {Solar} {Dynamics} {Observatory}}.{\BBCQ}
\newblock

\PrintBackRefs{\CurrentBib}

\bibitem [\protect \citeauthoryear {%
De Pontieu%
\ \protect \BOthers {.}}{%
De Pontieu%
\ \protect \BOthers {.}}{%
{\protect \APACyear {2014}}%
}]{%
depontieu_interface_2014_IRIS}
\APACinsertmetastar {%
depontieu_interface_2014_IRIS}%
\begin{APACrefauthors}%
De Pontieu, B.%
, Title, A\BPBI M.%
, Lemen, J\BPBI R.%
, Kushner, G\BPBI D.%
, Akin, D\BPBI J.%
, Allard, B.%
\BDBL {}Waltham, N.%
\end{APACrefauthors}%
\unskip\
\newblock
\APACrefYearMonthDay{2014}{{\APACmonth{07}}}{}.
\newblock
{\BBOQ}\APACrefatitle {The {Interface} {Region} {Imaging} {Spectrograph} ({IRIS})} {The {Interface} {Region} {Imaging} {Spectrograph} ({IRIS})}.{\BBCQ}
\newblock
\APACjournalVolNumPages{Solar Physics}{289}{7}{2733--2779}.
\newblock
\begin{APACrefURL} [{2023-03-28}]\url{http://link.springer.com/10.1007/s11207-014-0485-y} \end{APACrefURL}
\newblock
\begin{APACrefDOI} \doi{10.1007/s11207-014-0485-y} \end{APACrefDOI}
\PrintBackRefs{\CurrentBib}

\bibitem [\protect \citeauthoryear {%
Grubecka%
\ \protect \BOthers {.}}{%
Grubecka%
\ \protect \BOthers {.}}{%
{\protect \APACyear {2016}}%
}]{%
grubecka_height_2016}
\APACinsertmetastar {%
grubecka_height_2016}%
\begin{APACrefauthors}%
Grubecka, M.%
, Schmieder, B.%
, Berlicki, A.%
, Heinzel, P.%
, Dalmasse, K.%
\BCBL {}\ \BBA {} Mein, P.%
\end{APACrefauthors}%
\unskip\
\newblock
\APACrefYearMonthDay{2016}{{\APACmonth{09}}}{}.
\newblock
{\BBOQ}\APACrefatitle {Height formation of bright points observed by {IRIS} in {Mg} {II} line wings during flux emergence} {Height formation of bright points observed by {IRIS} in {Mg} {II} line wings during flux emergence}.{\BBCQ}
\newblock
\APACjournalVolNumPages{Astronomy \& Astrophysics}{593}{}{A32}.
\newblock
\begin{APACrefURL} [{2023-11-06}]\url{http://www.aanda.org/10.1051/0004-6361/201527358} \end{APACrefURL}
\newblock
\begin{APACrefDOI} \doi{10.1051/0004-6361/201527358} \end{APACrefDOI}
\PrintBackRefs{\CurrentBib}

\bibitem [\protect \citeauthoryear {%
{Henriques}%
\ \protect \BOthers {.}}{%
{Henriques}%
\ \protect \BOthers {.}}{%
{\protect \APACyear {2022}}%
}]{%
SALSA_guia}
\APACinsertmetastar {%
SALSA_guia}%
\begin{APACrefauthors}%
{Henriques}, V\BPBI M\BPBI J.%
, {Jafarzadeh}, S.%
, {Guevara G{\'o}mez}, J\BPBI C.%
, {Eklund}, H.%
, {Wedemeyer}, S.%
, {Szydlarski}, M.%
\BDBL {}{Mohan}, A.%
\end{APACrefauthors}%
\unskip\
\newblock
\APACrefYearMonthDay{2022}{{\APACmonth{03}}}{}.
\newblock
{\BBOQ}\APACrefatitle {{The Solar ALMA Science Archive (SALSA). First release, SALAT, and FITS header standard}} {{The Solar ALMA Science Archive (SALSA). First release, SALAT, and FITS header standard}}.{\BBCQ}
\newblock
\APACjournalVolNumPages{\aap}{659}{}{A31}.
\newblock
\begin{APACrefDOI} \doi{10.1051/0004-6361/202142291} \end{APACrefDOI}
\PrintBackRefs{\CurrentBib}

\bibitem [\protect \citeauthoryear {%
Lemen%
\ \protect \BOthers {.}}{%
Lemen%
\ \protect \BOthers {.}}{%
{\protect \APACyear {2012}}%
}]{%
lemen_atmospheric_2012}
\APACinsertmetastar {%
lemen_atmospheric_2012}%
\begin{APACrefauthors}%
Lemen, J\BPBI R.%
, Title, A\BPBI M.%
, Akin, D\BPBI J.%
, Boerner, P\BPBI F.%
, Chou, C.%
, Drake, J\BPBI F.%
\BDBL {}Waltham, N.%
\end{APACrefauthors}%
\unskip\
\newblock
\APACrefYearMonthDay{2012}{{\APACmonth{01}}}{}.
\newblock
{\BBOQ}\APACrefatitle {The {Atmospheric} {Imaging} {Assembly} ({AIA}) on the {Solar} {Dynamics} {Observatory} ({SDO})} {The {Atmospheric} {Imaging} {Assembly} ({AIA}) on the {Solar} {Dynamics} {Observatory} ({SDO})}.{\BBCQ}
\newblock
\APACjournalVolNumPages{Solar Physics}{275}{1-2}{17--40}.
\newblock
\begin{APACrefURL} [{2023-03-28}]\url{https://link.springer.com/10.1007/s11207-011-9776-8} \end{APACrefURL}
\newblock
\begin{APACrefDOI} \doi{10.1007/s11207-011-9776-8} \end{APACrefDOI}
\PrintBackRefs{\CurrentBib}

\bibitem [\protect \citeauthoryear {%
Ordoñez~Araujo%
}{%
Ordoñez~Araujo%
}{%
{\protect \APACyear {2024}}%
}]{%
araujo_dynamics_2024}
\APACinsertmetastar {%
araujo_dynamics_2024}%
\begin{APACrefauthors}%
Ordoñez~Araujo, F\BPBI J.%
\end{APACrefauthors}%
\unskip\
\newblock
\APACrefYearMonthDay{2024}{}{}.
\newblock
{\BBOQ}\APACrefatitle {Dynamics of the {Cold} {Solar} {Chromosphere} {Observed} with {ALMA}} {Dynamics of the {Cold} {Solar} {Chromosphere} {Observed} with {ALMA}}.{\BBCQ}
\newblock
\APACjournalVolNumPages{National University of Colombia, Repository}{}{}{}.
\newblock
\begin{APACrefDOI} \doi{https://repositorio.unal.edu.co/handle/unal/85838} \end{APACrefDOI}
\PrintBackRefs{\CurrentBib}

\bibitem [\protect \citeauthoryear {%
Ordoñez~Araujo%
, Guevara~Gómez%
\BCBL {}\ \BBA {} Calvo~Mozo%
}{%
Ordoñez~Araujo%
\ \protect \BOthers {.}}{%
{\protect \APACyear {2023}}%
}]{%
gibhub_reporsitory_alignment}
\APACinsertmetastar {%
gibhub_reporsitory_alignment}%
\begin{APACrefauthors}%
Ordoñez~Araujo, F\BPBI J.%
, Guevara~Gómez, J\BPBI C.%
\BCBL {}\ \BBA {} Calvo~Mozo, B.%
\end{APACrefauthors}%
\unskip\
\newblock
\APACrefYearMonthDay{2023}{{\APACmonth{04}}}{}.
\newblock
\APACrefbtitle {Alignment-of-IRIS-ALMA-and-SDO-images.} {Alignment-of-iris-alma-and-sdo-images.}
\newblock
\APAChowpublished {\url{https://github.com/JavierOrdonezA/Alignment-of-IRIS-ALMA-and-SDO-images.git}}.
\newblock
\APACrefnote{Accessed on April 13th, 2023.}
\PrintBackRefs{\CurrentBib}

\bibitem [\protect \citeauthoryear {%
{Pesnell}%
, {Thompson}%
\BCBL {}\ \BBA {} {Chamberlin}%
}{%
{Pesnell}%
\ \protect \BOthers {.}}{%
{\protect \APACyear {2012}}%
}]{%
SDO_pesnell}
\APACinsertmetastar {%
SDO_pesnell}%
\begin{APACrefauthors}%
{Pesnell}, W\BPBI D.%
, {Thompson}, B\BPBI J.%
\BCBL {}\ \BBA {} {Chamberlin}, P\BPBI C.%
\end{APACrefauthors}%
\unskip\
\newblock
\APACrefYearMonthDay{2012}{{\APACmonth{01}}}{}.
\newblock
{\BBOQ}\APACrefatitle {{The Solar Dynamics Observatory (SDO)}} {{The Solar Dynamics Observatory (SDO)}}.{\BBCQ}
\newblock
\APACjournalVolNumPages{\solphys}{275}{1-2}{3-15}.
\newblock
\begin{APACrefDOI} \doi{10.1007/s11207-011-9841-3} \end{APACrefDOI}
\PrintBackRefs{\CurrentBib}

\bibitem [\protect \citeauthoryear {%
Scherrer%
\ \protect \BOthers {.}}{%
Scherrer%
\ \protect \BOthers {.}}{%
{\protect \APACyear {2012}}%
}]{%
scherrer_helioseismic_2012}
\APACinsertmetastar {%
scherrer_helioseismic_2012}%
\begin{APACrefauthors}%
Scherrer, P\BPBI H.%
, Schou, J.%
, Bush, R\BPBI I.%
, Kosovichev, A\BPBI G.%
, Bogart, R\BPBI S.%
, Hoeksema, J\BPBI T.%
\BDBL {}Tomczyk, S.%
\end{APACrefauthors}%
\unskip\
\newblock
\APACrefYearMonthDay{2012}{{\APACmonth{01}}}{}.
\newblock
{\BBOQ}\APACrefatitle {The {Helioseismic} and {Magnetic} {Imager} ({HMI}) {Investigation} for the {Solar} {Dynamics} {Observatory} ({SDO})} {The {Helioseismic} and {Magnetic} {Imager} ({HMI}) {Investigation} for the {Solar} {Dynamics} {Observatory} ({SDO})}.{\BBCQ}
\newblock
\APACjournalVolNumPages{Solar Physics}{275}{1-2}{207--227}.
\newblock
\begin{APACrefURL} [{2022-02-09}]\url{http://link.springer.com/10.1007/s11207-011-9834-2} \end{APACrefURL}
\newblock
\begin{APACrefDOI} \doi{10.1007/s11207-011-9834-2} \end{APACrefDOI}
\PrintBackRefs{\CurrentBib}

\bibitem [\protect \citeauthoryear {%
Starovoitov%
, E.E.%
\BCBL {}\ \BBA {} K.T.%
}{%
Starovoitov%
\ \protect \BOthers {.}}{%
{\protect \APACyear {2020}}%
}]{%
sim_vs_pcc_2020}
\APACinsertmetastar {%
sim_vs_pcc_2020}%
\begin{APACrefauthors}%
Starovoitov, V.%
, E.E., E.%
\BCBL {}\ \BBA {} K.T., I.%
\end{APACrefauthors}%
\unskip\
\newblock
\APACrefYearMonthDay{2020}{01}{}.
\newblock
{\BBOQ}\APACrefatitle {Comparative analysis of the SSIM index and the Pearson coefficient as a criterion for image similarity} {Comparative analysis of the ssim index and the pearson coefficient as a criterion for image similarity}.{\BBCQ}
\newblock
\APACjournalVolNumPages{Eurasian Journal of Mathematical and Computer Applications}{8}{}{76-90}.
\newblock
\begin{APACrefDOI} \doi{10.32523/2306-6172-2020-8-1-76-90} \end{APACrefDOI}
\PrintBackRefs{\CurrentBib}

\bibitem [\protect \citeauthoryear {%
Vernazza%
, Avrett%
\BCBL {}\ \BBA {} Loeser%
}{%
Vernazza%
\ \protect \BOthers {.}}{%
{\protect \APACyear {1981}}%
}]{%
vernazza_structure_1981}
\APACinsertmetastar {%
vernazza_structure_1981}%
\begin{APACrefauthors}%
Vernazza, J\BPBI E.%
, Avrett, E\BPBI H.%
\BCBL {}\ \BBA {} Loeser, R.%
\end{APACrefauthors}%
\unskip\
\newblock
\APACrefYearMonthDay{1981}{{\APACmonth{04}}}{}.
\newblock
{\BBOQ}\APACrefatitle {Structure of the solar chromosphere. {III} - {Models} of the {EUV} brightness components of the quiet-sun} {Structure of the solar chromosphere. {III} - {Models} of the {EUV} brightness components of the quiet-sun}.{\BBCQ}
\newblock
\APACjournalVolNumPages{The Astrophysical Journal Supplement Series}{45}{}{635}.
\newblock
\begin{APACrefURL} [{2023-05-11}]\url{http://adsabs.harvard.edu/doi/10.1086/190731} \end{APACrefURL}
\newblock
\begin{APACrefDOI} \doi{10.1086/190731} \end{APACrefDOI}
\PrintBackRefs{\CurrentBib}

\bibitem [\protect \citeauthoryear {%
Wang%
, Bovik%
, Sheikh%
\BCBL {}\ \BBA {} Simoncelli%
}{%
Wang%
\ \protect \BOthers {.}}{%
{\protect \APACyear {2004}}%
}]{%
wang_image_2004_SSIM}
\APACinsertmetastar {%
wang_image_2004_SSIM}%
\begin{APACrefauthors}%
Wang, Z.%
, Bovik, A.%
, Sheikh, H.%
\BCBL {}\ \BBA {} Simoncelli, E.%
\end{APACrefauthors}%
\unskip\
\newblock
\APACrefYearMonthDay{2004}{{\APACmonth{04}}}{}.
\newblock
{\BBOQ}\APACrefatitle {Image {Quality} {Assessment}: {From} {Error} {Visibility} to {Structural} {Similarity}} {Image {Quality} {Assessment}: {From} {Error} {Visibility} to {Structural} {Similarity}}.{\BBCQ}
\newblock
\APACjournalVolNumPages{IEEE Transactions on Image Processing}{13}{4}{600--612}.
\newblock
\begin{APACrefURL} [{2023-03-29}]\url{http://ieeexplore.ieee.org/document/1284395/} \end{APACrefURL}
\newblock
\begin{APACrefDOI} \doi{10.1109/TIP.2003.819861} \end{APACrefDOI}
\PrintBackRefs{\CurrentBib}

\bibitem [\protect \citeauthoryear {%
Wedemeyer%
\ \protect \BOthers {.}}{%
Wedemeyer%
\ \protect \BOthers {.}}{%
{\protect \APACyear {2020}}%
}]{%
wedemeyer_sun_2020_ALMA_band_3}
\APACinsertmetastar {%
wedemeyer_sun_2020_ALMA_band_3}%
\begin{APACrefauthors}%
Wedemeyer, S.%
, Szydlarski, M.%
, Jafarzadeh, S.%
, Eklund, H.%
, Guevara~Gomez, J\BPBI C.%
, Bastian, T.%
\BDBL {}Carlsson, M.%
\end{APACrefauthors}%
\unskip\
\newblock
\APACrefYearMonthDay{2020}{{\APACmonth{03}}}{}.
\newblock
{\BBOQ}\APACrefatitle {The {Sun} at millimeter wavelengths: {I}. {Introduction} to {ALMA} {Band} 3 observations} {The {Sun} at millimeter wavelengths: {I}. {Introduction} to {ALMA} {Band} 3 observations}.{\BBCQ}
\newblock
\APACjournalVolNumPages{Astronomy \& Astrophysics}{635}{}{A71}.
\newblock
\begin{APACrefURL} [{2023-04-17}]\url{https://www.aanda.org/10.1051/0004-6361/201937122} \end{APACrefURL}
\newblock
\begin{APACrefDOI} \doi{10.1051/0004-6361/201937122} \end{APACrefDOI}
\PrintBackRefs{\CurrentBib}

\bibitem [\protect \citeauthoryear {%
Wootten%
\ \BBA {} Thompson%
}{%
Wootten%
\ \BBA {} Thompson%
}{%
{\protect \APACyear {2009}}%
}]{%
wootten_atacama_2009}
\APACinsertmetastar {%
wootten_atacama_2009}%
\begin{APACrefauthors}%
Wootten, A.%
\BCBT {}\ \BBA {} Thompson, A\BPBI R.%
\end{APACrefauthors}%
\unskip\
\newblock
\APACrefYearMonthDay{2009}{{\APACmonth{08}}}{}.
\newblock
{\BBOQ}\APACrefatitle {The {Atacama} {Large} {Millimeter}/submillimeter {Array}} {The {Atacama} {Large} {Millimeter}/submillimeter {Array}}.{\BBCQ}
\newblock
\APACjournalVolNumPages{Proceedings of the IEEE}{97}{8}{1463--1471}.
\newblock
\begin{APACrefURL} [{2023-11-27}]\url{http://arxiv.org/abs/0904.3739} \end{APACrefURL}
\newblock
\APACrefnote{arXiv:0904.3739 [astro-ph]}
\newblock
\begin{APACrefDOI} \doi{10.1109/JPROC.2009.2020572} \end{APACrefDOI}
\PrintBackRefs{\CurrentBib}

\bibitem [\protect \citeauthoryear {%
Yeager%
}{%
Yeager%
}{%
{\protect \APACyear {2023}}%
}]{%
yeager_libguides_nodate}
\APACinsertmetastar {%
yeager_libguides_nodate}%
\begin{APACrefauthors}%
Yeager, K.%
\end{APACrefauthors}%
\unskip\
\newblock
\APACrefYearMonthDay{2023}{{\APACmonth{03}}}{}.
\newblock
\APACrefbtitle {{LibGuides}: {SPSS} {Tutorials}: {Pearson} {Correlation}.} {{LibGuides}: {SPSS} {Tutorials}: {Pearson} {Correlation}.}
\newblock
\begin{APACrefURL} [{2023-03-29}]\url{https://libguides.library.kent.edu/SPSS/PearsonCorr} \end{APACrefURL}
\PrintBackRefs{\CurrentBib}

\end{thebibliography}

\end{document}